\documentclass[aps, prl, superscriptaddress, twocolumn, amsfonts, amsmath, amssymb, floatfix]{revtex4-1}

\usepackage{graphicx}
\usepackage{dcolumn}
\usepackage{bm}
\usepackage{epsfig}
\usepackage{hyperref}
\usepackage{braket}
\usepackage{color}
\usepackage{amsmath}

\begin{document}

\title{Observation of Spin Superfluidity in a Bose Gas Mixture}

\author{Eleonora Fava}
\affiliation{INO-CNR BEC Center and Dipartimento di Fisica, Universit\`a di Trento, 38123 Povo, Italy}

\author{Tom Bienaim\'e}
\affiliation{INO-CNR BEC Center and Dipartimento di Fisica, Universit\`a di Trento, 38123 Povo, Italy}

\author{Carmelo Mordini}
\affiliation{INO-CNR BEC Center and Dipartimento di Fisica, Universit\`a di Trento, 38123 Povo, Italy}
\affiliation{Trento Institute for Fundamental Physics and Applications, INFN, 38123 Povo, Italy}

\author{Giacomo Colzi}
\affiliation{INO-CNR BEC Center and Dipartimento di Fisica, Universit\`a di Trento, 38123 Povo, Italy}
\affiliation{Trento Institute for Fundamental Physics and Applications, INFN, 38123 Povo, Italy}

\author{Chunlei Qu}
\affiliation{INO-CNR BEC Center and Dipartimento di Fisica, Universit\`a di Trento, 38123 Povo, Italy}

\author{Sandro Stringari}
\affiliation{INO-CNR BEC Center and Dipartimento di Fisica, Universit\`a di Trento, 38123 Povo, Italy}
\affiliation{Trento Institute for Fundamental Physics and Applications, INFN, 38123 Povo, Italy}

\author{Giacomo Lamporesi}
\affiliation{INO-CNR BEC Center and Dipartimento di Fisica, Universit\`a di Trento, 38123 Povo, Italy}
\affiliation{Trento Institute for Fundamental Physics and Applications, INFN, 38123 Povo, Italy}

\author{Gabriele Ferrari}
\affiliation{INO-CNR BEC Center and Dipartimento di Fisica, Universit\`a di Trento, 38123 Povo, Italy}
\affiliation{Trento Institute for Fundamental Physics and Applications, INFN, 38123 Povo, Italy}

\date{\today}

\begin{abstract}

The spin dynamics of a harmonically trapped Bose--Einstein condensed binary mixture of sodium atoms is experimentally investigated at finite temperature. In the collisional regime the motion of the thermal component is shown to be damped because of spin drag, while the two condensates exhibit a counterflow oscillation without friction, thereby providing direct evidence for spin superfluidity.  Results are also reported in the collisionless regime where the spin components of both the condensate and thermal part oscillate without damping, their relative motion being driven by a mean-field effect. We also measure the static polarizability of the condensed and thermal parts and we find a large increase of the condensate  polarizability with respect to the T=0 value, in agreement with the predictions of theory.

\end{abstract}

\maketitle

In the last years, spin-superfluidity and spin-transport phenomena have attracted a great interest in the community of condensed matter physics from both the experimental and theoretical point of view \cite{Sonin10}. Even in systems where spin is conserved, the behavior of spin transport is highly nontrivial since, at finite temperature, collisions between different spin species yield relaxation of the spin current, a phenomenon known as spin drag. So far the study of superfluidity at finite temperature has mainly concerned the density channel, where both the number of particles and total current are conserved. A major consequence is that, in the collisional regime, sound can propagate both in the superfluid phase, where it takes the form of first and second sound, as well as in the normal phase (ordinary sound). In the presence of collisions, spin sound can instead propagate only in the superfluid phase, so that its observation, in this case, can be considered as an ultimate proof of spin superfluidity. In fact, the propagation of spin sound in the collisionless regime is consistent with superfluidity, but can be predicted also in the normal phase as a consequence of mean-field interactions (see, for instance, the propagation of sound in a normal Fermi liquid \cite{Pines66}).  
Actually the equations of hydrodynamics applied to a superfluid quantum mixture predict the propagation of three sounds \cite{Volovik75,Andreev75}: pressure, temperature and spin sound  (see \cite{Armaitis15} for a recent application of three-velocity hydrodynamic theory to Bose--Bose mixtures). 

The dynamic behavior of multicomponent quantum gases has been extensively investigated  in the last years (see, for example, \cite{Stamper-Kurn13} for a review on spinor Bose gases). Experiments on spin dynamics have been carried out in gases occupying two different hyperfine states,   \cite{Hall98,Maddaloni00,DeMarco02,Nascimbene09,Hamner11,Nicklas11,Sidorenkov13,Bienaime16,Valtolina17}, in larger spinor systems \cite{Stenger98,Schmaljohann04,Higbie05,Pagano14,Zibold16,Kim17}, as well as in mixtures of different isotopes or atomic species \cite{Modugno00,Modugno02,Ferlaino03,Ferrier-Barbut14}. Theoretical activity in these systems has also become very popular  (see, for example, \cite{Vichi99,Jezek02,Rodriguez04,Mur-Petit06,Zhang07,Recati11,Sartori13,Zeng14,Armaitis15,Abad15,Castin15,Lee16,Armaitis17}). Spin-drag phenomena have been experimentally investigated in the unitary Fermi gas \cite{Joseph15,Sommer11,Bardon14}, in Bose gases \cite{Koller15}, in Bose--Fermi mixtures \cite{Delehaye15}, as well as in two-dimensional Fermi gases \cite{Koschorreck13,Luciuk17}. The role of spin polarization on the stability of supercurrents \cite{Beattie13} and the counterflow instability in Bose--Fermi \cite{Ferrier-Barbut14} and in Bose--Bose \cite{Hoefer11,Kim17} mixtures have also been experimentally investigated. 

In this Letter, we experimentally study the  spin-dipole oscillation and the role of collisions at finite temperature. The main result of our  work is the observation of undamped spin oscillations in the collisional regime. This observation actually provides direct evidence of spin superfluidity.

We consider a symmetric BEC mixture of the $|m_F=+1\rangle \equiv \ket{\uparrow}$ and $|m_F=-1\rangle \equiv \ket{\downarrow}$ components of the $F=1$ hyperfine ground state of sodium atoms, confined in a harmonic trap. Differently from most of the quantum mixtures so far investigated in the literature, our sodium mixture is characterized by an almost perfect symmetry between the two components, both in terms of the number of atoms occupying the two hyperfine states, the confining potential and the intraspecies interaction. Furthermore the mixture is fully-miscible, not subject to buoyancy and is close to the miscible-immiscible phase transition since $(a-a_{\uparrow\downarrow})/a = 0.07 \ll 1$, with $a \equiv a_{\uparrow\uparrow}=a_{\downarrow\downarrow}=54.54(20) a_0$ and $a_{\uparrow\downarrow}=50.78(40)a_0$  \cite{Knoop11}, $a_0$ being the Bohr radius. This mixture, then, represents an ideal system to investigate the effects of spin superfluidity.  The zero temperature behavior of the spin-dipole oscillation was investigated in \cite{Bienaime16}.   Here, we report results at finite temperature, both in the collisional  and in the collisionless regimes, which are experimentally realized by varying the frequencies of the trapping potential.  We prove that in both regimes the mixture is able to support undamped spin oscillations. Furthermore, the vicinity to the miscible-immiscible phase transition is associated with a strong coupling between the two spin clouds. In addition to the softening of the spin-dipole oscillation frequency and the sizable increase of the static spin polarizability, that were already observed at zero temperature \cite{Bienaime16}, the vicinity to the phase transition  causes a further important amplification of the spin polarization of the superfluid component due to the interaction with the thermal part.

We start with an equally populated mixture of the $\uparrow,\downarrow$ states \cite{Bienaime16,SM} with $N_\uparrow = N_\downarrow \simeq 4\times \, 10^5$ (with a spin imbalance fluctuation smaller than 10\%) and consider two different trap geometries: (A) a crossed optical trap with frequencies $\left[ \omega_x,\omega_y,\omega_z \right] / 2 \pi = \left[87,330,250\right] \, \text{Hz}$  and (B) a single-beam optical trap with frequencies $\left[ \omega_x,\omega_y,\omega_z \right] / 2 \pi = \left[12,1350,1350  \right] \, \text{Hz}$. Using parametric heating, we can adjust the condensed fraction of the mixture, i.e., the ratio between the total number of atoms in the condensates $N_0$ and the total number of atoms $N=N_{\uparrow}+N_{\downarrow}$. A major difference between the two configurations is that, in the long axial direction, configuration (A) is basically characterized by a collisionless regime ($\omega_x \tau_{\uparrow\downarrow}\gg 1$), while configuration (B) by a more collisional one ($\omega_x \tau_{\uparrow\downarrow} \sim 1$). The difference is not due to significant changes in the density, but rather in the value of $\omega_x$. The collisional time between the $\uparrow,\downarrow$ components can be estimated employing the classical expression for $\tau_{\uparrow\downarrow}$, with the density calculated in the center of the trap at $T=T_c$ \cite{SM}. We estimate $\omega_x \tau_{\uparrow\downarrow}$ of a few tens in configuration (A) and of order unity in configuration (B).\\

\begin{figure}[t!]
\centerline{{\includegraphics[width=\columnwidth]{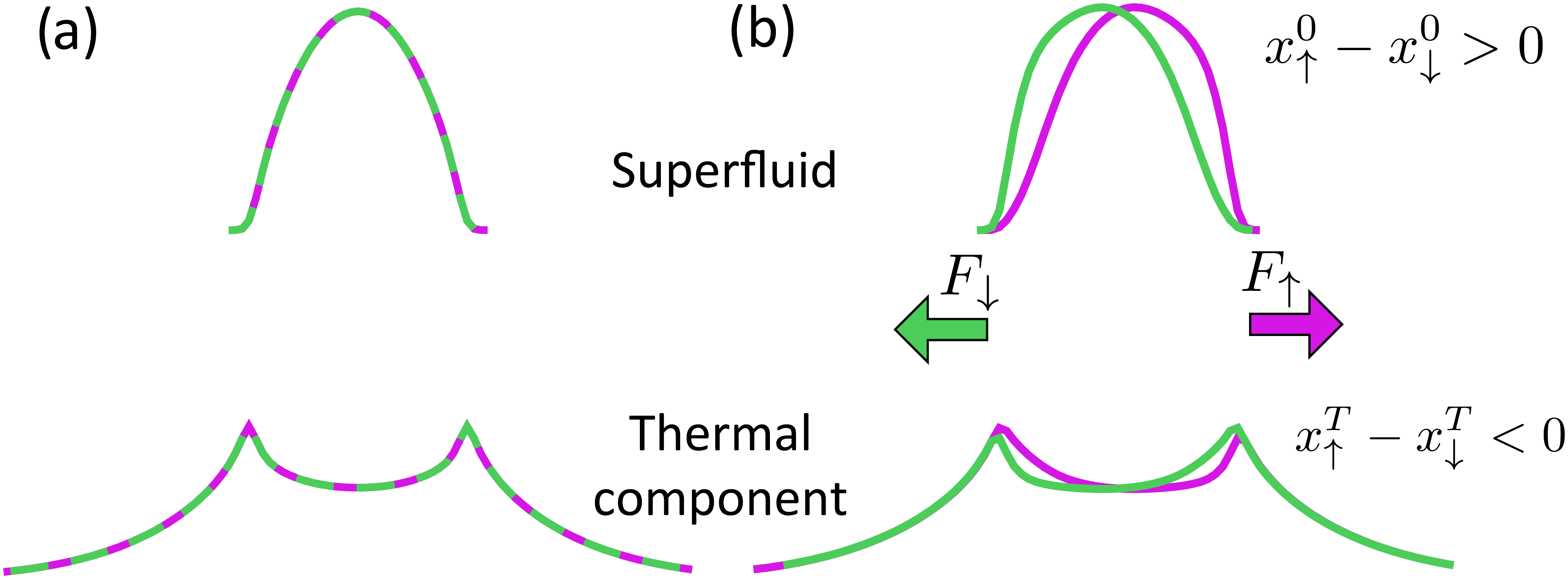}}}
\caption{Computed atomic density distribution $n_{\uparrow,\downarrow}(x,0,0)$ of the binary mixture at finite temperature showing the component $\uparrow$ (violet) and $\downarrow$ (green), each one of these being composed of a superfluid (top) and a thermal part (bottom). (a) In the absence of any external force the centers of mass of all four components overlap. (b) In the presence of a differential force $F_{\uparrow,\downarrow}$, the condensed part shows a large positive polarization, while the thermal component interacting with the condensate is polarized in the opposite direction. The thermal part lying outside the BEC region has a small positive polarization.} \label{Fig1}
\end{figure}

{\em Spin dynamics}. The spin oscillation is excited by applying a magnetic field gradient $B'_x$ for a few ms. This generates a small force $F_{\uparrow,\downarrow} = \pm g_F \mu_B B'_x$ ($g_F$ is the Land\'e factor, $\mu_B$ the Bohr magneton) that tends to separate the two components, as illustrated in Fig. \ref{Fig1}. Such a procedure leaves the total center of mass at rest and gives rise to time-varying spin displacements
\begin{align*}
    S_0 &\equiv x^{0}_\uparrow - x^{0}_\downarrow, \qquad   S_T \equiv x^{T}_\uparrow - x^{T}_\downarrow,
\end{align*}
of both the condensed $S_0$ and the thermal part $S_T$, where $x^{0}_{\uparrow,\downarrow}$  and $x^{T}_{\uparrow,\downarrow}$ are the centers of the atomic distribution of the condensed and thermal components of the $\uparrow,\downarrow$ density distributions. In the experiment, we are able to study the dynamics of such four-fluid system by monitoring 
each of the four components to reconstruct $S_0$ and  $S_T$ as a function of time. The amplitude of oscillation of $\left\{S_0,S_T \right\}$ is smaller than the Thomas--Fermi radius $R_x$ of the cloud [for a fully Bose--Einstein condensed mixture at $T=0$, $R_x = 25 \, \mu\text{m}$ for configuration (A) and $R_x = 230 \, \mu\text{m}$ for  (B)].
The two spin states are separately imaged after a Stern--Gerlach expansion in a magnetic field gradient along $z$, which allows us to extract the centers of mass of the four components of the fluid $\left\{ x^{0}_\uparrow,x^{0}_\downarrow,x^{T}_\uparrow,x^{T}_\downarrow\right\}$ \cite{SM}.

\begin{figure*}[!t]
\centerline{{\includegraphics[width=0.9\textwidth]{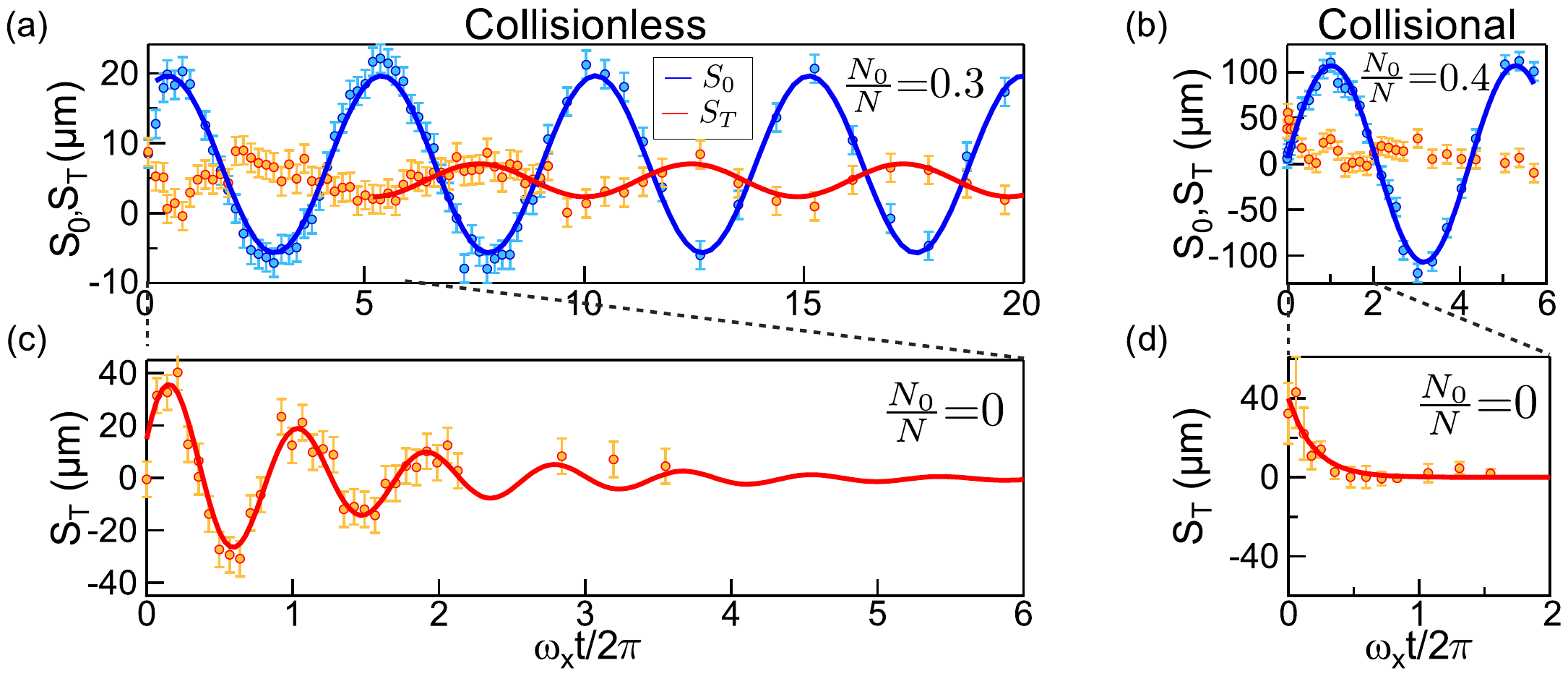}}}
\caption{(a) Spin oscillations for the thermal $S_T$ (red) and condensed $S_0$ (blue) parts of the mixture with $N_0/N = 0.3$ ($T/T_c\simeq0.85$) for configuration (A). After a small transient period, $S_T$ oscillates at $\omega_T =0.207(2) \, \omega_x$ which turns out to be equal, within errorbars, to the oscillation frequency of $S_0$, $\omega_{\text{SD}} = 0.205(2) \, \omega_x$. The ratio of the oscillation amplitude of $S_T$ and $S_0$ is $0.18(2)$. (b) Spin oscillations for the condensed and the thermal $\left\{S_0,S_T \right\}$ parts for a mixture with $N_0/N = 0.4$ ($T/T_c \simeq 0.75$) in configuration (B). The condensed component oscillates at $\omega_{\text{SD}} = 0.233(5) \, \omega_x$, while the thermal relative motion is quickly damped. We measure an exponential decay of $S_T$ corresponding to $\omega_x \tau = 1.5(6)$. (c) Thermal spin current $S_T$ for a non-superfluid mixture (above $T_c$) in configuration (A) where we observe a few damped oscillations at the trap frequency $\omega_x$ with an exponentially decaying envelope from which we extract the decay lifetime, and obtain $\omega_x \tau = 11(2)$. (d) Same measurement for configuration (B) where we observe a purely exponential decay and extract $\omega_x \tau = 1.2(4)$, compatible with the measurement of $\tau$ below $T_c$. To maintain a roughly constant condensed fraction during the measurement, we limit the observation time to the first $500 \, \text{ms}$ after excitation. This explains why, due to the different trapping frequency $\omega_x$, more oscillations are shown for configuration (A) than for (B).} \label{Fig2}
\end{figure*}

The spin dynamics of the condensate is shown in Fig. \ref{Fig2}a and \ref{Fig2}b at relatively high values of $T/T_c$, corresponding to  $N_0/N\sim 0.3$ and $N_0/N\sim 0.4$, respectively. The figure shows that the condensate, in the presence of a large thermal component, exhibits spin oscillations without visible damping in both collisionless (A) and collisional (B) regimes. 
The absence of friction near the BEC border, where the Landau critical velocity is vanishingly small, is due to the fact that the spin velocity, during the spin-dipole oscillation, is strongly suppressed near the surface of the condensate (see Fig.\ref{1}b), differently from what happens in the rigid motion of the center-of-mass oscillation, and in agreement with the Steinwedel--Jensen model for the isospin oscillations of nuclear physics \cite{Bohr98}. 
The measured frequencies ($\omega_{\text{SD}} = 0.205(2) \, \omega_x$ in (A) and $\omega_{\text{SD}} = 0.233(5) \, \omega_x$ in (B)) differ by about 6\% from the value reported in Ref.~\cite{Bienaime16} at very low temperatures ($\omega_{\text{SD}} = 0.218(2)\, \omega_x$) and by 7\% (A) and 20\% (B) from the value $\omega_{\text{SD}}^0=\sqrt{(a-a_{\uparrow\downarrow})/(a+a_{\uparrow\downarrow})} \,\omega_x = 0.19(2) \, \omega_x$ \cite{Sartori15}, predicted by hydrodynamic theory of superfluids at $T=0$ 
 \footnote{ The hydrodynamic result $\omega_{\text{SD}}^0$ for the collective frequency is independent of the number of atoms and of the density of the condensate. It holds in the limit of small amplitude oscillations and in the Thomas--Fermi approximation which, in the case of spin oscillations, requires the condition that the spin healing length $\xi_s= 1/ \sqrt{8 \pi n (a-a_{\uparrow\downarrow})}$ be much smaller than the Thomas--Fermi radius \cite{Sartori15}. The softening of the frequency for values of the scattering lengths close to the demixing transition is consistent with the softening of the spin sound velocity in uniform matter given, at $T=0$, by the expression $c_s = \sqrt{ \frac{n}{2m} \frac{4 \pi \hbar^2}{m} (a-a_{\uparrow\downarrow}) }$.}.

The thermal component, instead, behaves very differently in the two regimes. In the collisionless regime (A), after a transient of damped oscillations, it oscillates at the same spin-dipole frequency of the condensate, but with opposite phase and a smaller amplitude (see Fig. \ref{Fig2}a), the ratio between the thermal and the condensed amplitudes being $0.18(2)$.  In the collisional regime (B), the thermal part is instead strongly damped and quickly reaches an equilibrium position, where both spin thermal components are at rest in the center of the trap (see Fig.~\ref{Fig2}b) 
\footnote{Actually, we observe a small residual oscillations of the thermal part in opposite phase with respect to the condensed component (the ratio of the oscillation amplitudes of $S_T$ and $S_0$ is smaller than $0.04(1)$) which is expected to disappear in the deep collisional regime $\omega_x \tau_{\uparrow\downarrow} \ll 1$.}. 

In Fig. \ref{Fig2}, we report the results for spin dynamics above $T_c$, as well. In configuration (A) the cloud exhibits several oscillations before relaxing, thus revealing that collisions are not very strong (Fig. \ref{Fig2}c). \emph{Viceversa}, in the collisional regime (B), the behavior is diffusive, suggesting an overdamped spin oscillation (Fig. \ref{Fig2}d). A similar spin-drag effect was observed in the Bose-Fermi mixture of \cite{Delehaye15}, as well as in a Bose gas above $T_c$ in \cite{Koller15}. From our experimental data, we extract $\omega_x \tau = 11(2)$ for (A) and $\omega_x \tau = 1.2(4)$ for (B). These measurements are in agreement with the theoretical estimates of $\omega_x \tau_{\uparrow\downarrow}$ given earlier in the Letter.

Finally, it is worth pointing out that the behavior of the spin-dipole oscillations is very different with respect to the center-of-mass motion, where both the condensed and thermal parts oscillate in phase without damping at the frequency $\omega_x/2\pi$ \cite{SM}, independent of the presence of collisions.

{\em Spin-dipole polarizability}. The counter-phase oscillation of the thermal component observed in the collisionless regime (see Fig. \ref{Fig2}a) can be physically understood by investigating the behavior of  the spin-dipole polarizability of the gas at finite temperature, employing the mean field Hartree--Fock theory \cite{Pitaevskii16} in the presence of a static spin-dipole constraint of the form $-m \omega_x^2 x_0 x\sigma_z$ ($\sigma_z$ is the third Pauli matrix). This additional potential generates a force acting on the two spin components  in opposite directions ($F_{\uparrow,\downarrow}= \pm m\omega^2_x x_0$),  $x_0$ being  the displacement of the trap minimum for each component. By neglecting  interaction effects induced by the thermal component on the condensate, as well as thermal-thermal interactions, and using the Thomas--Fermi approximation for the condensate, one obtains the following result for the spin density $s_z^{0} =n^{0}_\uparrow-n^0_\downarrow$ of the condensate \cite{Sartori15}:
\begin{equation}
s_z^{0} = -x_0 \frac{a+a_{\uparrow \downarrow}}  {a-a_{\uparrow \downarrow}}\frac{ \partial n^0}{\partial x} \, .
\label{1}
\end{equation}
For the spin density $s_z^{T} =n^T_\uparrow-n^T_\downarrow$  of the thermal component one instead finds the results
\begin{equation}
s_z^{T} = -x_0 \frac{a+a_{\uparrow \downarrow}}{a-a_{\uparrow \downarrow}}\frac{ \partial n^T}{\partial x}\label{2}
\end{equation}
inside the spatial region occupied by the condensate, where the thermal part feels interaction  effects,  and 
\begin{equation}
s_z^{T} = -x_0\frac{ \partial n^T}{\partial x}
\label{2}
\end{equation}
outside.  In the above equations, $n^{0}$ and $n^T$ are the equilibrium condensate and thermal total densities, respectively. The corresponding contribution to the spin-dipole polarizability is then obtained  by integrating the quantities $xs_z^0$ and $xs_z^T$. 
These results show that the spin polarization of  the inner thermal atoms (see Eq. (2))  is amplified by the same large  factor  $(a+a_{\uparrow\downarrow})/(a-a_{\uparrow\downarrow})$  as for the  condensate.  The corresponding polarization effects have  however opposite signs, the density derivative of the condensate, at equilibrium, being opposite to the one of the inside thermal component (see Fig. \ref{Fig1}).

\begin{figure}[t]
\centerline{{\includegraphics[width=\columnwidth]{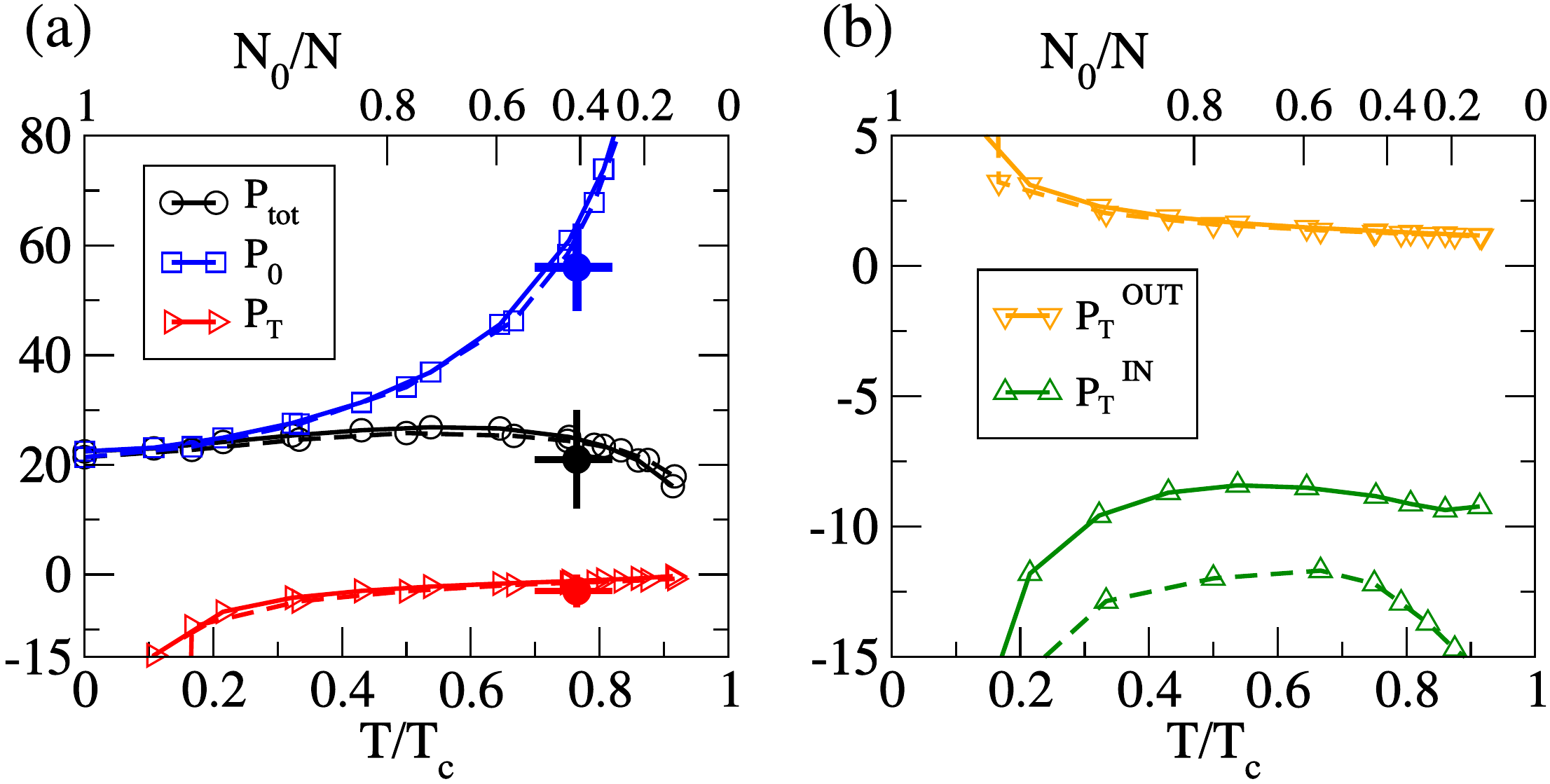}}}
\caption{ (a) Static spin-dipole polarizability as a function of temperature showing, respectively, the different contributions from the superfluid (blue), the thermal component (red) and the total one (black). (b) The thermal part lying in the region occupied by the superfluid has a negative polarization (green) whereas the outer part has a small positive polarization.
The calculation has been performed for the two different configurations (A) (solid) and (B) (dashed). The static polarizabilities measured for $N_0/N=0.4$ are also shown and well agrees with the predictions of theory.} \label{Fig3}
\end{figure}

For higher temperatures, interaction effects of the thermal component on the condensate can no longer be neglected. The behavior of the spin polarization can be explored more accurately, by solving in a consistent way the coupled Hartree--Fock equations for the condensate and for the thermal part \cite{SM}. Figure \ref{Fig3} shows the resulting predictions for the condensate and thermal contributions to the spin polarizability, which are respectively defined as $\mathcal P_0 = (\int x s_z^{0} d \mathbf{r})/N_0$ and  $\mathcal P_T = (\int x s_z^{T} d \mathbf{r})/N_T$. The figure reveals the occurrence of a large enhancement of $\mathcal P_0$ with respect to the $T=0$ case, which is caused by the interaction with the inside thermal component and is strongly enhanced by the smallness of $(a-a_{\uparrow\downarrow})$. The resulting values for the temperature dependence of the polarization of the condensate, as well as of the total polarization, $\mathcal{P}_\mathrm{tot}=(N_0 \mathcal{P}_0+N_T \mathcal{P}_T)/N$, turn out to be practically the same in the regimes (A) and (B) considered in this work. Despite the large increase of $\mathcal P_0$, the  total polarization $\mathcal{P}_\mathrm{tot}$ turns out to be practically independent of $T$ in a wide range of temperatures.
The above discussion suggests that, in the collisionless regime, the thermal atoms are locked to the condensate and oscillate in opposite phase in the spin-dipole dynamics. 
In the collisional regime (Fig.~\ref{Fig2}b), instead, the thermal part quickly relaxes to equilibrium, because of spin drag.

\begin{figure}[t]
\centerline{{\includegraphics[width=\columnwidth]{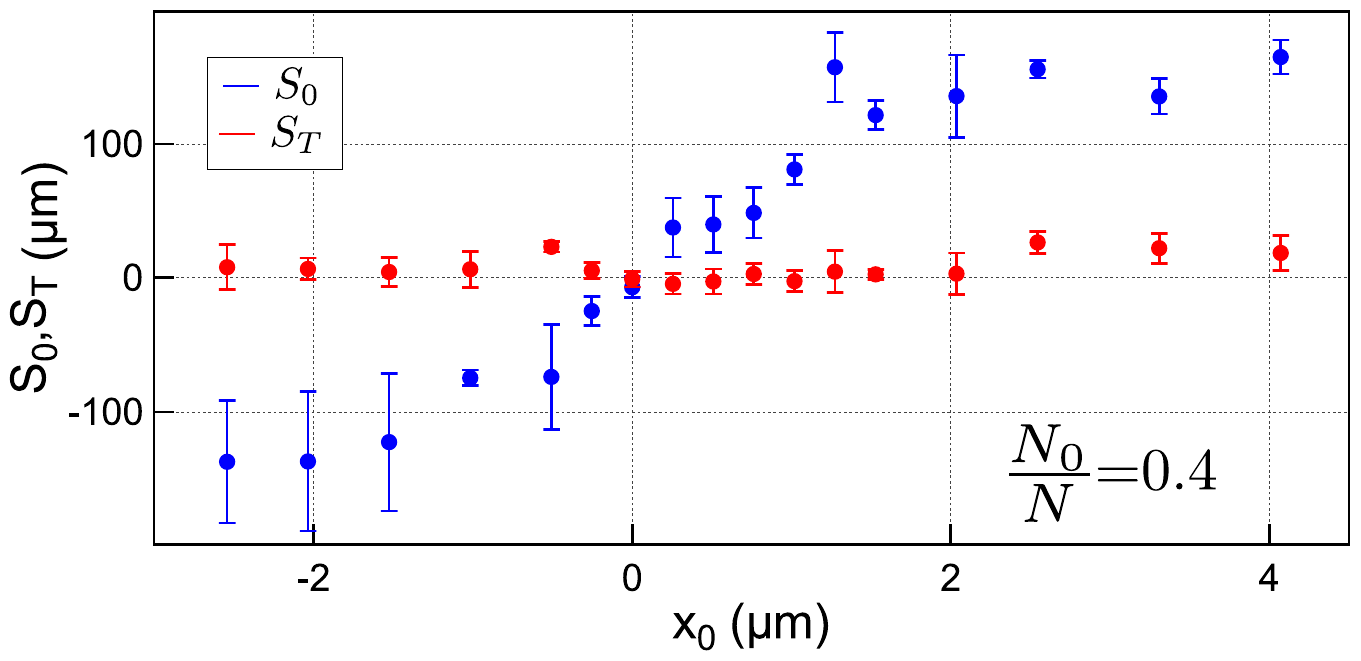}}}
\caption{Measured spin displacements $\left\{S_0,S_T\right\}$ for the thermal (red) and condensed components (blue) of the mixture as a function of $x_0$ for $N_0/N = 0.4$ for configuration (B). From such data, we extract $\left\{\mathcal P_0,\mathcal P_T \right\}$ using a linear fit to the data in the linear region around the origin. We obtain $\left\{\mathcal P_0=56(8),\mathcal P_T=-3(3) \right\}$. } \label{Fig4}
\end{figure}

Using the experimental method introduced in Ref.~\cite{Bienaime16} we measure the static spin polarizability for the trap geometry (B) and identify the contributions that arise from the condensate and from the thermal part. Starting with both $\uparrow,\downarrow$ components perfectly overlapped in the harmonic potential, we apply a slowly increasing force $F_{\uparrow,\downarrow}$ to each component that eventually shifts their trap minima by $\pm x_0$. 
In this way the global center of mass is unaffected, while the superfluid and thermal spin components acquire finite relative displacements $\left\{S_0,S_T \right\}$. The spin polarizability of the condensed and thermal fractions $\left\{  \mathcal P_0 \equiv S_0/(2 x_0), \mathcal P_T \equiv S_T/(2 x_0)      \right\}$ are extracted in the linear regime, \emph{i.e.}, for values of $\left\{S_0,S_T\right\}$ much smaller than the Thomas--Fermi radius of the condensed component \cite{Bienaime16,Sartori15}.  Figure \ref{Fig4} shows the spin displacements $\left\{S_0,S_T\right\}$ of the thermal and condensed components of the mixture as a function of $x_0$ for $N_0/N =  0.4$. From this data, we extract the polarizability by performing a linear fit around the origin. The region where we fit the data to extract the value of the polarizabilities corresponds to the small $x_0$ linear regime ($R_x = 230 \, \mu\text{m}$ is the Thomas--Fermi radius along $x$). The analysis of the data points out the occurrence of a large polarization of the condensate, in accordance with the predictions of theory (see Fig. \ref{Fig3}). \\

In conclusion, we have investigated the spin dynamics and the spin polarizability of a superfluid Bose--Bose mixture at finite temperature. Our results reveal the occurrence of undamped  spin oscillations, which are observed  not only in the collisionless regime, where the mean field drives a counter-phase oscillation of the thermal part, but also in the presence of strong collisions, which are responsible for the relaxation of the thermal component, because of spin drag. The absence of friction of the spin motion in the collisional regime provides a direct proof of the spin superfluid nature of the system.  We have also shown that, thanks to the vicinity to the miscible-immiscible phase transition,  the   interaction   between the two spin clouds causes, at finite temperature,  a large increase of the polarizability of the condensate with respect to the $T=0$ value. Natural generalizations of the present work concern the study of persistent spin currents in ring geometries and the propagation of spin sound waves and magnetic solitons \cite{Qu16}.

\begin{acknowledgments}
We thank L. Pitaevskii, A. Recati, C. Fort, F. Minardi, N. Proukakis and E. Sonin for useful comments and discussions. We acknowledge funding by  the  Provincia  Autonoma  di  Trento, the QUIC grant of the Horizon 2020 FET program, and by the Istituto Nazionale di Fisica Nucleare.
\end{acknowledgments}

\newpage
.
\newpage


\section*{Supplemental material}

\noindent\textit{Experimental procedure}\\

In the experiment, we use a symmetric mixture of the $|F,m_F\rangle = |1,+1 \rangle \equiv \ket{\uparrow}$ and $|1,-1 \rangle \equiv \ket{\downarrow}$ states of ultra-cold $^{23}$Na atoms that are confined either in (A) a crossed optical dipole trap with frequencies $\left[ \omega_x,\omega_y,\omega_z \right] / 2 \pi = \left[87,330,250\right] \, \text{Hz}$ or in (B) a single-beam optical trap with frequencies $\left[ \omega_x,\omega_y,\omega_z \right] / 2 \pi = \left[12,1350,1350  \right] \, \text{Hz}$. The magnetic fields along the three spatial directions are calibrated with a precision of $1 \, \text{mG}$ using RF spectroscopy techniques.
We start with a fully polarized Bose--Einstein condensate in $\ket{\downarrow}$. The first step towards the creation of the spin mixture is to perform a Landau--Zener transition to the $|1,0\rangle$ state with nearly $100 \, \%$ transfer efficiency.
This is realized at a magnetic field of $100 \, \text{G}$ to isolate a two-level system exploiting the quadratic Zeeman shifts. The second step consists in inducing a Rabi oscillation among the three Zeeman sublevels to obtain a 50/50 spin mixture of $\ket{\downarrow}$ and $\ket{\uparrow}$ \cite{Bienaime16,Zibold16}. 
The bias field along $\hat x$ is taken small enough to allow us to neglect the quadratic Zeeman shifts compared to the Rabi frequency and is kept on during the whole experimental sequence following the Rabi pulse. The number of atoms in each spin component is $N_\uparrow = N_\downarrow \simeq 4\times \, 10^5$. In order to prevent the decay of the mixture to $|1,0\rangle$ by spin changing collisions, we lift this level by $\sim h\times 1 \, \text{kHz}$ using blue detuned microwave dressing on the transition to $|2,0\rangle$. In this way, we deal with a stable two-component system $\ket{\uparrow}$ and  $\ket{\downarrow}$ .

The sample is heated using parametric excitation at twice the trapping frequency along the radial direction. We adjust the number of cycles of the warm-up procedure (at fixed amplitude) to control the temperature between $650 \, \text{nK}$ and $1 \, \text{$\mu$K}$, while the condensed fraction varies from $40 \, \%$ to $0 \, \%$ (fully thermal sample). An important experimental aspect is that the dipole trap should be deep enough to prevent any evaporation during the heating process. We check this point by monitoring the total atom number after the heating procedure and verify that it is conserved. To distinguish between the center of the atomic distribution of the condensed ($x^0_{\uparrow,\downarrow}$) and thermal ($x^T_{\uparrow,\downarrow}$) part we use a bimodal fitting function with independent centers for the two distributions. Taking into account the integration along the imaging probe direction and the time of flight, the resulting distributions are well described by a Gaussian and a Thomas--Fermi function for the thermal and condensed density profiles respectively. All the images are taken performing a Stern--Gerlach expansion for a few ms ($3$ ms for configuration (A) and $7.5$ ms for (B)) which does not affect the correct estimation of $x^0_{\uparrow,\downarrow}$ and $x^T_{\uparrow,\downarrow}$. 

\begin{figure}[t]
\centerline{{\includegraphics[width=0.5\textwidth]{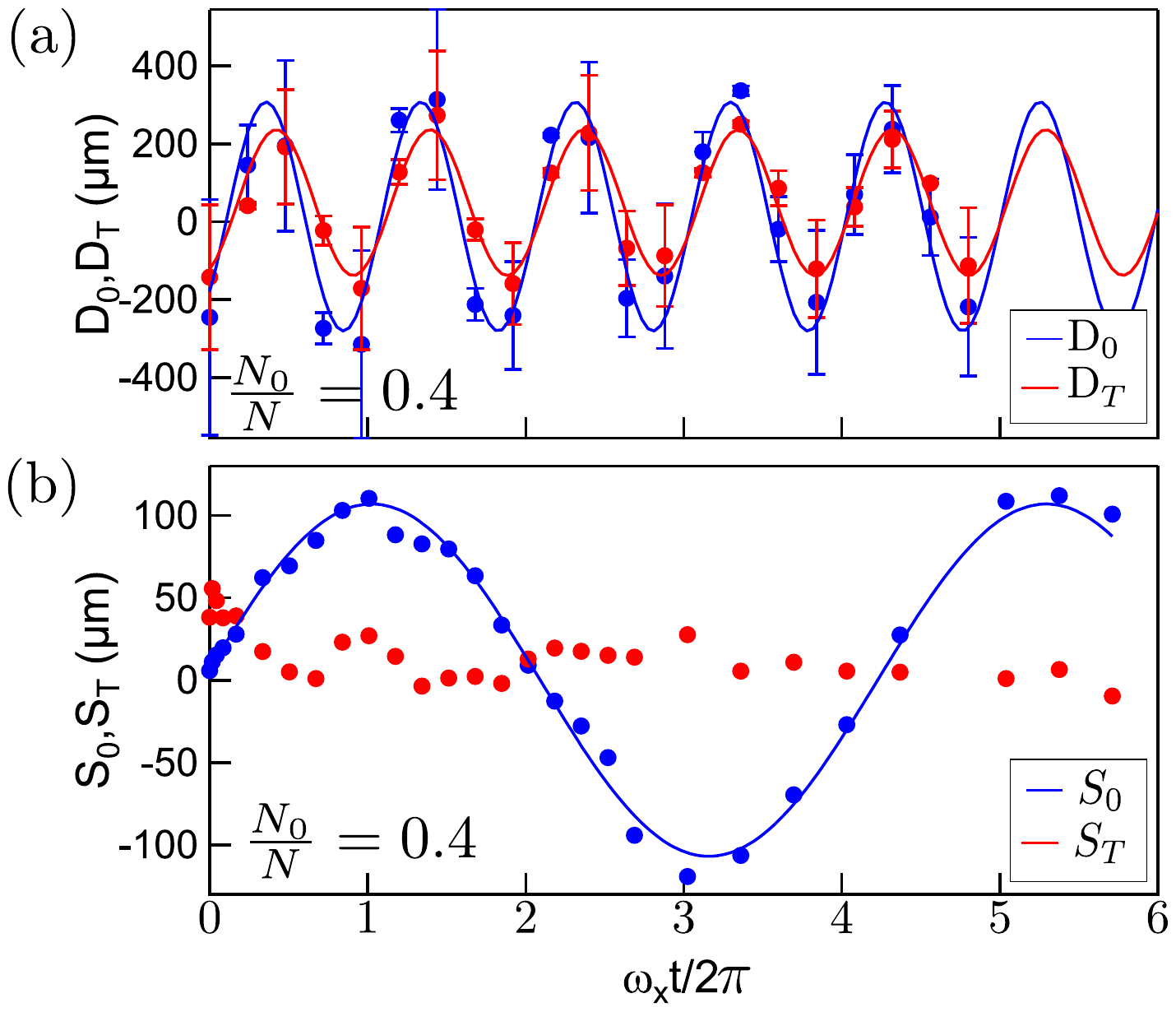}}}
\caption{Dynamics of (a) the center of mass and of (b) the spin dipole current for the thermal and condensed components in configuration (B) with $N_0/N=0.4$.} \label{FigS1}
\end{figure}

In a complementary experiment, we excite and monitor as a function of time the center of mass of the condensed and thermal part
\begin{align*}
    D_0 &\equiv x^{0}_\uparrow + x^{0}_\downarrow, \qquad   D_T \equiv x^{T}_\uparrow + x^{T}_\downarrow \, ,
\end{align*}
which are related to the density current. The data of this experiment carried out in configuration (B), are shown in Fig. \ref{FigS1} where we also report for comparison the oscillation of the spin-dipole mode already reported in Fig. \ref{Fig2} of the Letter. The density current presents a completely different behavior with respect to the spin dynamics, being undamped for both the condensed and thermal fraction in the most collisional regime (B) as reported in Fig. \ref{FigS1}.

The collisional time $\tau_{\uparrow\downarrow}$ between the thermal $\uparrow$ and $\downarrow$ components is evaluated using the following expression:
\begin{equation*}
1/\tau_{\uparrow\downarrow}=n^T_{\uparrow\downarrow} \sigma_{\uparrow\downarrow} v_{\text{rel}},
\end{equation*}
where $n^T_{\uparrow\downarrow} = \left[ m \bar \omega^2/(2 \pi k_B) \right]^{3/2} N_{\uparrow\downarrow} T^{-3/2}$ is the peak density of the thermal distribution, $\sigma_{\uparrow\downarrow} = 4 \pi a_{\uparrow\downarrow}^2$ is the cross section of two distinguishable particles and $v_{\text{rel}}=\sqrt{8 k_B T/(\pi m_r)}$ is their average relative velocity ($N_{\uparrow\downarrow}=N/2$, $m_r=m/2$ is the reduced mass and $\bar \omega = (\omega_x\omega_y\omega_z)^{1/3}$).\\

\noindent\textit{Sum rules and spin-dipole oscillation frequency}\\

A useful estimate of the spin-dipole frequency is provided by the ratio
\begin{equation}
\hbar\omega_{\text{SD}}=\sqrt{\frac{m_1(S_D)}{m_{-1}(S_D)}} \, ,
\label{eq:sumrule}
\end{equation}
between the energy weighted and inverse energy weighted moment of the spin-dipole  dynamical structure factor
\begin{equation}
S_{\text{SD}}(\omega)=Q^{-1}\sum_{m,n}e^{-E_n/k_BT}|\langle n |S_D|m \rangle|^2\delta(\hbar \omega-\hbar \omega_{nm})
\label{Somega}
\end{equation}
where $S_D= \sum_k x_k\sigma_z^k$ is the spin dipole operator, $\omega_{nm}=(E_n-E_m)/\hbar$ are the Bohr frequencies and $Q=\sum_m e^{-E_m/k_BT}$ is the usual partition function.
The energy weighted moment is model independent and given by the Thomas-Reich-Kuhn sum rule $m_1=N\hbar^2/2m$ \cite{Pitaevskii16}.  The inverse energy weighted moment is instead related to the dimensionless static spin-dipole polarizability $\mathcal P_{\text{tot}}$, fixed by the ratio between the induced spin displacement  of the atomic cloud  and  the separation $2x_0$ of the two harmonic traps, through the equation \cite{Sartori15}
\begin{equation}
m_{-1}(S_D )=\frac{N\mathcal P_{\text{tot}}}{2m\omega_x^2} \; .
\end{equation}
In the absence of interaction between opposite spins, one has $\mathcal P_{\text{tot}}=1$.
As discussed in the main tex, the linear polarizability can  be calculated carrying out a static Hartree--Fock calculation in the presence of a small external static constraint,  proportional to the spin-dipole operator. The calculation can be done both at $T=0$, where it corresponds to the solution of the Gross-Pitaevskii equation, and at finite temperature  (see next  section)
At zero  temperature the resulting estimate 
 \begin{equation}
\omega_{\text{SD}}=\omega_x /\sqrt{\mathcal P_{\text{tot}}}
\label{eq:sumruleP}
\end{equation}
for the spin-dipole frequency  turns out to be very accurate because the collective state practically exhausts the sum rules $m_1$ and $m_{-1}$, the dynamic structure factor being  characterized by a single peak. This was well confirmed by the experiment of \cite{Bienaime16}. The experimental results of Fig. \ref{Fig2} suggest that also at finite temperature the excitation spectrum is basically characterized by a single collective frequency. So it is not a surprise that the ratio  (\ref{eq:sumruleP}) provides a reasonable estimate of the observed collective frequency also at finite  temperature.\\

\noindent\textit{Static spin-dipole polarizability}\\

The condensed order parameters $\Psi_j$ ($j=\uparrow,\downarrow$), in the presence of the spin displacement of the external potentials  $V_{\uparrow}=V_{ho}(x-x_0,y,z)$ and   $V_{\downarrow}=V_{ho}(x+x_0,y,z)$, satisfy, at equilibrium, the following  Gross-Pitaevskii equations:
\begin{equation}
\mu\Psi_j=\left[-\frac{\hbar^2}{2m}\nabla^2+U^0_j
\right]\Psi_j
\end{equation}
where $U^0_j$ is the effective potential felt by the condensate of the $j$-th component
\begin{eqnarray}
U^0_{\uparrow}&=&V_{\uparrow}+g(n^0_{\uparrow}+2n^T_{\uparrow})+g_{\uparrow\downarrow}(n^0_{\downarrow}+n^T_{\downarrow}) \label{psi1}\\
U^0_{\downarrow}&=&V_{\downarrow}+g(n^0_{\downarrow}+2n^T_{\downarrow})+g_{\uparrow\downarrow}(n^0_{\uparrow}+n^T_{\uparrow}) \label{psi2}
\end{eqnarray}
with $g=4\pi \hbar^2 a/m$ and $g_{\uparrow\downarrow}=4\pi \hbar^2 a_{\uparrow\downarrow}/m$. The condensate densities are given by $n^0_j=|\Psi_j|^2$ and we have assumed $N_\uparrow=N_\downarrow$.
The thermal atom densities are determined by the semi-classical equation
\begin{equation}
n^T_{j}(\mathbf{r})=\frac{1}{(2\pi\hbar)^3}\int d\mathbf{p} f_j(\mathbf{p},\mathbf{r},t) \;  ,
\label{eq:wigner}
\end{equation}
where the Wigner distribution function of the thermal atoms is given by
\begin{equation}
f_j(\mathbf{p},\mathbf{r},t)=\{e^{\beta[\mathbf{p}^2/2m+U^T_{j}-\mu]}-1\}^{-1} \;  ,
\end{equation}
the potentials for the thermal normal fluid are 
\begin{eqnarray}
U^T_{\uparrow} &=& V_{\uparrow} + 2g(n^0_{\uparrow}+n^T_{\uparrow})+g_{\uparrow \downarrow}(n^0_{\downarrow}+n^T_{\downarrow}) \\
U^T_{\downarrow} &=& V_{\downarrow} + 2g(n^T_{\downarrow}+n^T_{\downarrow})+g_{\uparrow \downarrow}(n^0_{\uparrow}+n^T_{\uparrow})
\end{eqnarray}

In terms of the Bose function
\begin{equation}
g_{3/2}(z)=\frac{2}{\sqrt{\pi}}\int_0^\infty dx \frac{\sqrt{x}}{z^{-1}e^{x}-1}
\end{equation}
Eq.~(\ref{eq:wigner}) then reduces to
\begin{eqnarray}
n^T_{j} &=& g_{3/2}(z_j)/\lambda^3_T
\end{eqnarray}
where 
$\lambda_T=\sqrt{2\pi\hbar^2/(mk_BT)}$ is the thermal De Broglie wavelength and 
\begin{eqnarray}
z_j &=& \exp[(\mu-U^T_{j})/k_BT] 
\end{eqnarray}
is the local fugacity of each spin component.
We solve the above  coupled equations to find the ground state density distributions for the condensate and thermal atoms in the presence of the displacement $2x_0$ of the two harmonic traps and thus obtain the polarizability of the system as a function of $x_0$ and temperature $T$. The results for the linear polarizabilities of the condensate and of the thermal part are reported in Fig. \ref{Fig3} of the main text.

From the above equations it is straightforward to derive the results (1-3) reported in the main text of this Letter and holding at low temperature. To this purpose we  apply the Thomas-Fermi approximation for the condensate wavefunction and neglect the interaction between the condensate and the thermal component in the Gross-Pitaevskii equations (\ref{psi1},\ref{psi2}) for the condensate. This straightforwardly yields result (\ref{1}) for the spin density of the condensate $s_z^{0} =n^{0}_\uparrow-n^0_\downarrow$. For the thermal part one instead finds the result
\begin{equation*}
s_z^T =n^{T}_\uparrow-n^T_\downarrow = -\frac{1}{k_BT\lambda^3_T}z \frac{\partial g_{3/2}(z)}{\partial z} \frac{2g}{g-g_{\uparrow \downarrow}}m\omega^2_xx_0 \;  .
\end{equation*}
in the inside region where the thermal part interacts with the condensate.  On the other hand,  in the absence of the spin-dipole external perturbation, one finds 
\begin{equation*}
\frac{\partial n^T}{\partial x} = \frac{1}{k_BT\lambda^3_T}z\frac{\partial g_{3/2}(z)}{\partial z} \frac{2g}{g+g_{\uparrow \downarrow}} m\omega^2_xx_0 \;  ,
\end{equation*}
so that Eq. (2) is immediately recovered. In analogous way one derives result (3) for the spin polarization of the outermost thermal component.\\

\end{document}